\documentclass[prb,reprint,a4paper,floatfix,superscriptaddress]{revtex4-1}
\usepackage{graphicx}
\usepackage{physics}
\usepackage{hyperref}

\newcommand{\etal}{\textit{et al. }}

\begin{document}

\title{\boldmath
The probability of an encounter of photons in nested and double-nested Mach-Zehnder interferometers
	\unboldmath	}

	\author{E. Schmidt}
	\email{schmidt@physics.muni.cz}
	\affiliation{Faculty of Science, Masaryk University, Kotl\'{a}\v{r}sk\'a 2, 61137 Brno, Czech Republic}

	\author{A. Dubroka}
	\affiliation{Faculty of Science, Masaryk University, Kotl\'{a}\v{r}sk\'a 2, 61137 Brno, Czech Republic}
	
	\date{\today}
	
	\pacs{ }
	
\begin{abstract}
We present the results of a theoretical work discussing the propagation of an electromagnetic wave through nested Mach-Zehnder interferometers using classical optics and standard quantum theory. We show that some seemingly surprising effects at first sight, which are often explained in the literature using the two-state vector formalism (TSVF), are a direct consequence of destructive or constructive interference and thus there is no need for the unconventional TSVF formulation. We show that the probability of a photon detection derived from the weak value used in TSVF can be interpreted as the probability of an encounter of two opposing photon fluxes.
\end{abstract}

\maketitle

\section{Introduction}
The literature about photon behavior is very rich and extensive. One of the often recurring problems concerns the question ``where was the photon?" passing an optical system, e.g., an interferometer. Our paper is inspired by two experimental articles of Danan~\etal\cite{Danan2013} and the recently published article of Zhou~\etal\cite{Zhou} that explain their experiment using the two-state vector formalism (TSVF), which is based on the theoretical paper of L. Vaidman \cite{Vaidman}. 
The basic idea of the TSVF approach is that for a more complete description of a quantum system, one should be concerned not only with the forward-evolving wave function $\ket{\psi}$ but also the backward-evolving wave function $\bra{\varphi}$.
A knowledge of both should give some information about the system which can be tested by the so-called weak measurement, which should only weakly affect photon behavior. The publication of the TSVF approach was followed by a prompt response \cite{Li,VaidmanComment,Lundeen}  and a long series of concurring or critical opinions  \cite{Salih,Danan2015,Vaidman2014,Huang,Saldanha,Bartkiewicz2015,Potocek,Vaidman2016,Vaidman2016Comment,Bartkiewicz2016,Hashmi2016,Li2015,Ben-Israel,Alonso,
	Vaidman2018,Bula,Len,Englert,
	Sokolovski2016,Sokolovski2017,Griffiths2016,
	Nikolaev,Duprey,Vaidman2017Comment,Griffiths,
	VaidmanJETP2017,Nikolaev2017,Sokolovski2017Comment,Peleg,Englert2019}. 
In this paper, we discuss the propagation of photons in Mach-Zehnder interferometers (MZI) using classical optics and standard quantum-mechanical approach. We compare the results of this approach with the TSVF.

\begin{figure}
	\hspace*{-12mm}
	\includegraphics[width=9cm]{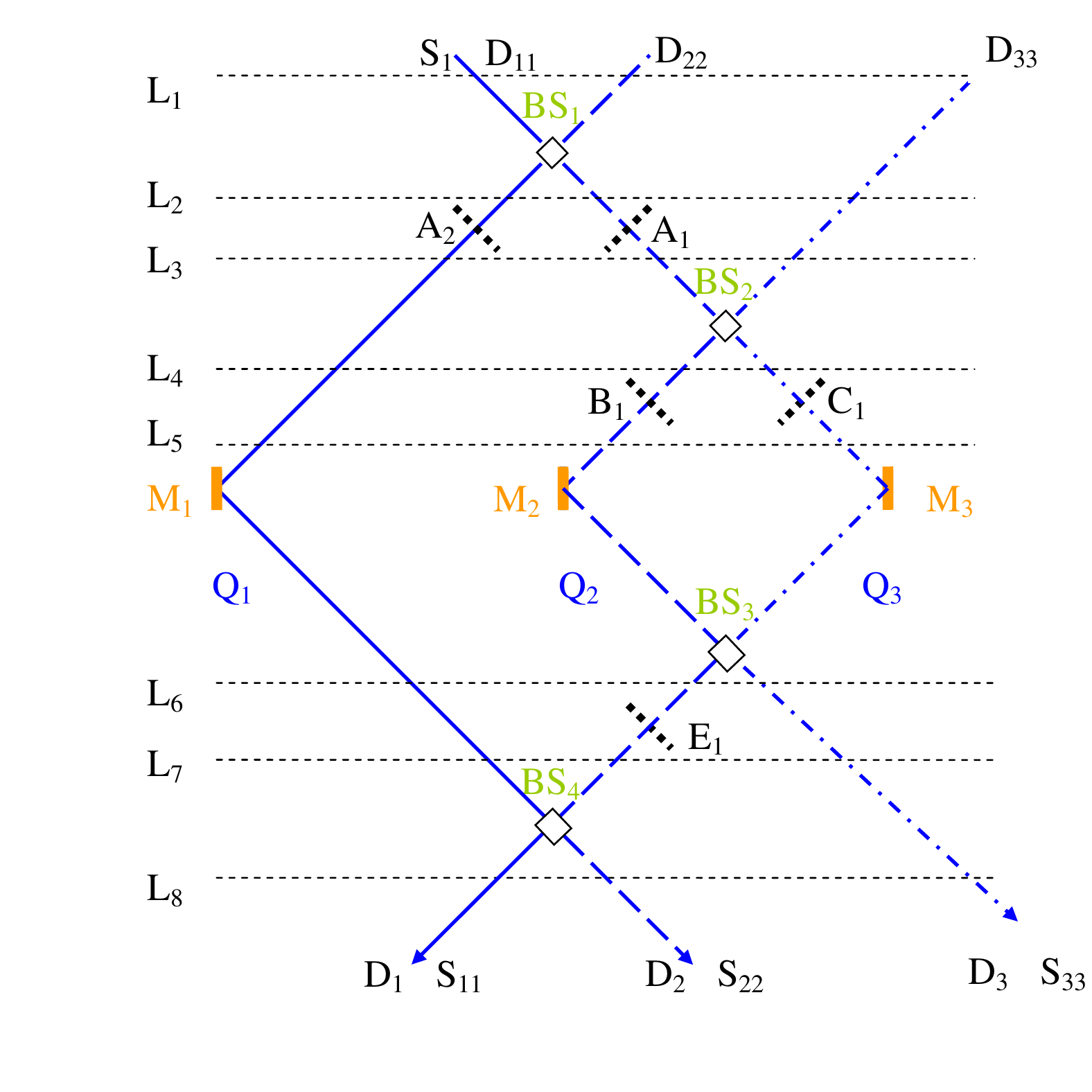}
	\vspace*{-7mm}
	\caption{\label{Fig1nMZscheme} The scheme of the single-nested Mach-Zehnder interferometer. S$_1$ and D$_i$ with $i=1, 2, 3$  (S$_{ii}$ and D$_{ii}$ with $ii=11, 22, 33$) denote sources and detectors, respectively, for the forward (backward) propagating beam. 	
	BS denote beam splitters, M denote mirrors and A$_1$, A$_2$, B$_1$, C$_1$, E$_1$ denote modulators. There are three optical paths (or quantum channels) Q$_k$ ($k=1,2,3$) shown by solid, dashed and dash-dotted blue lines, respectively. 	There are eight stages $L_m$ $(m=1\ldots8)$ shown by the horizontal dashed lines where we evaluate the forward and backward evolving wave functions and related intensities. 
	}
\end{figure}

We proceed along the lines of Li~\etal\cite{Li} and Hashmi~\etal~\cite{Hashmi2016} and consider the often used arrangement of the Mach-Zehnder interferometer (MZI) with one smaller nested MZI, see Fig.~\ref{Fig1nMZscheme}. For the sake of simplicity, we consider four identical ideal achromatic beam splitters BS with reflection coefficient $r$ and transmission coefficient $t$:\cite{Holbrow2002}
\begin{equation}
\label{r}
r=\frac{\rm i}{\sqrt{2}}\;,\quad t=\frac{\rm 1}{\sqrt{2}}\;.
\end{equation}
We consider three ideal mirrors M with reflectivity equal to 1, five modulators A$_1$, A$_2$, B$_1$, C$_1$, E$_1$ that can change the phase and amplitude of the wave function. We suppose that it is possible to measure the intensity of the beam propagating in the forward direction (from top to bottom in Fig.~\ref{Fig1nMZscheme}) and also in the backward direction (from bottom to top). For the forward direction, we consider one source of light S$_1$ and three detectors D$_1$, D$_2$, D$_3$. For the backward direction, we consider three sources S$_{11}$, S$_{22}$, S$_{33}$ and three detectors D$_{11}$, D$_{22}$, D$_{33}$. We consider three quantum channels (or optical paths) $Q_k$ ($k=1, 2, 3$) depicted in Fig.~1 by solid, dashed and dash-dotted lines, respectively,  and eight stages $L_m$  ($m=1\ldots8$) in order to distinguish the state before and after the modulation. We calculate the forward and backward evolving wave functions $\ket{\psi}$  and $\bra{\varphi}$, respectively, at the intersections of $Q_k$  and  $L_m$.

The experimental arrangement of Danan~\etal\cite{Danan2013} is very similar to the one presented in Fig.~\ref{Fig1nMZscheme}; it involves the source S$_1$ and the detector D$_2$. As modulators, vibrating mirrors with a specific frequency were used in positions equivalent to A$_1$, A$_2$, B$_1$, C$_1$, E$_1$ . 
We believe the difference in type of modulation is not substantial and should not affect the following discussion and conclusions. The main result of Danan~\etal\cite{Danan2013} is that the signal measured at the detector D$_2$ exhibited the frequency from the perturbation A$_2$, B$_1$, C$_1$ but not from A$_1$ and E$_1$. 
The authors of Ref.~\cite{Danan2013} interpreted the experiment using the TSVF and draw 
the nontrivial conclusion that photons come to the detector $D_2$ via  the channel $Q_1$ and at the same time are present inside the nested interferometer (in the sections with the modulators B$_1$ and  C$_1$) but not outside the nested interferometer (in the sections with the modulators A$_1$ and E$_1$). The latter results were associated with the conclusion that the past of the photons is not represented by continuous trajectories~\cite{Danan2013}. 

There are several other experimental works on MZI. For example, single photon variant of the experiment of Danan~\etal\cite{Danan2013} was realized by Zhou~\etal\cite{Zhou}. There exist also other variants, see Li~\etal\cite{Li2015}  and  Ben-Israel~\etal\cite{Ben-Israel}  who used a combination of a standard MZI with a nested MZI. Another modification of the nested MZI with Dove prism inside the nested part was tried by Alonso~\etal\cite{Alonso} followed by a comment~\cite{Vaidman2018}.
An alternative to the work of Danan~\etal\cite{Danan2013} was reported by Bula~\etal\cite{Bula} where a slightly modified spectral distribution of light was used for modulation instead of vibrating mirrors. The experiment of  Len~\etal\cite{Len} on the MZI using a single-photon source was recently published. Another variant of the single photon experiment using the MZI and nested MZI was suggested by Englert~\etal\cite{Englert}.

The main part of this paper presented in Sec.~\ref{nMZI}  is devoted to calculations of the beam intensity at different stages and quantum channels of the single nested MZI by means of classical optics and standard quantum theory. We also calculate the weak value, related probability and propose its  interpretation as the probability of an encounter of the forward and backward-propagating photon fluxes.  We show that the discontinuous patterns occurring in the weak value are direct consequence of destructive or constructive interference. In Sec.~\ref{nnMZI}, we shortly repeat the calculations for the symmetric double-nested MZI, and for completeness we consider the standard (simple) MZI in Sec.~\ref{simpleMZI}. In Sec.~\ref{perturbations}, we discuss the results for the nested MZI with modulators.

\section{Results}
\subsection{Single nested Mach-Zehnder interferometer}
\label{nMZI}
First we state our assumptions concerning light intensity and detection. We suppose that during the measurement, the photon is absorbed in an irreversible process and the detector is capable of measuring a large number of photons as well as individual photons. Formally we proceed in the same way as in the work of P.~L.~Saldanha~\cite{Saldanha}, i.e., we use a classical description of waves which is valid also for the wave function of photons. We assume that the beam is a plane wave, however, the results should be valid also for a Gaussian beam. We assume that the intensity of the beam is measurable at the output of the interferometer and at any stage inside the interferometer. Following Duarte~\cite{Duarte}, we assume that the beam of light is a high power laser beam consisting of a large number of photons. We adopt the Dirac's idea~\cite{Dirac} that each photon interferes with itself and assume that the results of such an experiment and the interpretation of results are independent of the number of photons, that is, they should be the same for an intensive flux of photons as well as for a single photon state. In the latter, the quantitative evaluation of the intensity should be treated statistically as a number of photons detected per a sufficiently long integration time. This should not be confused with the so-called ``weak measurement" or ``weak value" mentioned below. In this section, we treat the nonmodulated beam; the effects of modulation are discussed in Sec.~\ref{perturbations}.  

\begin{figure*}
	\vspace*{-5mm}
	\hspace*{-3mm}
	\includegraphics[width=18cm]{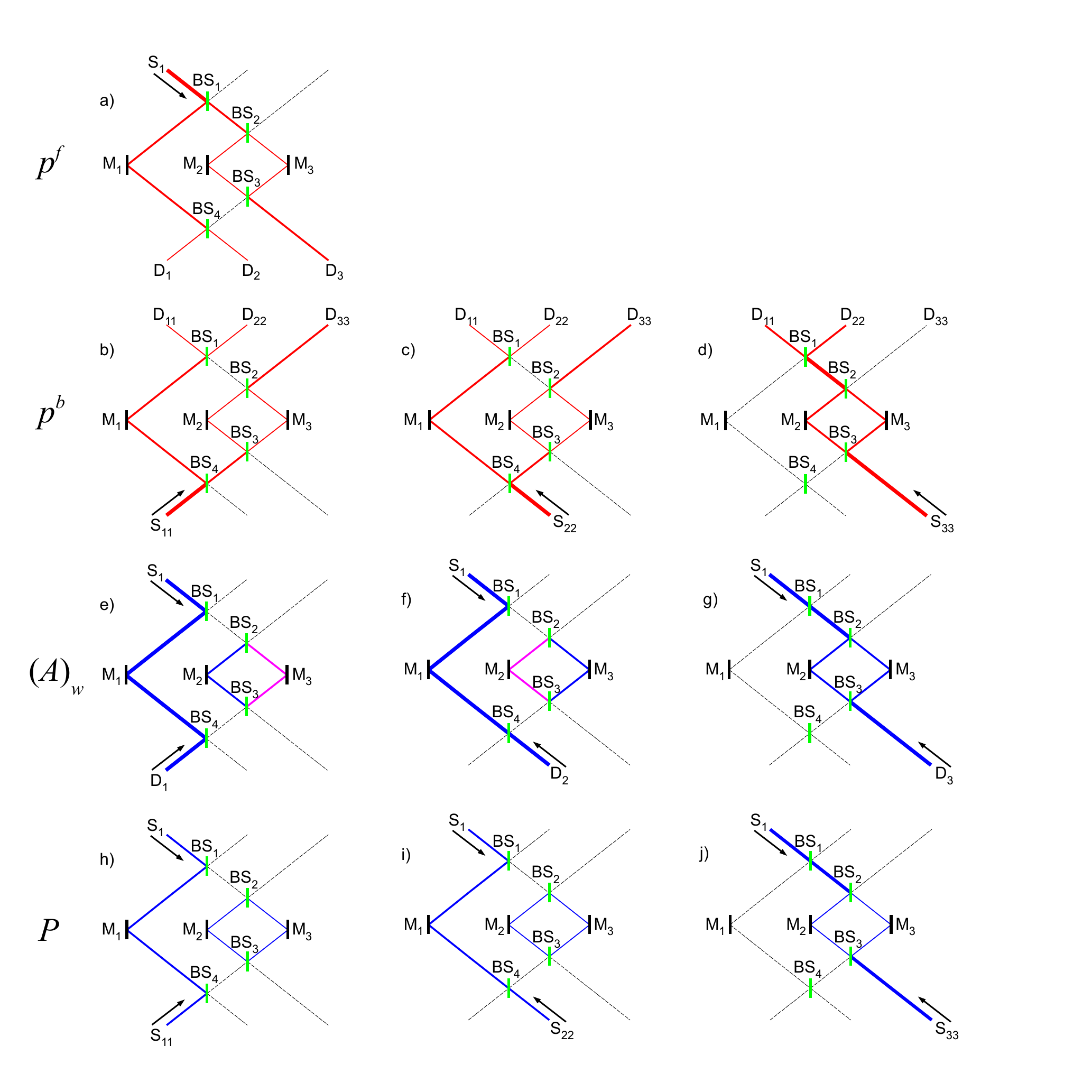}
	\vspace*{-10mm}
	\caption{\label{Fig2} Results for the nested Mach-Zehnder interferometer. M denotes a mirror  and BS a beam splitter (see also Fig. \ref{Fig1nMZscheme}). Panel (a) schematically shows intensities (or  probabilities $p^f$) of the forward-propagating beam from the source S$_1$ (red lines). Panels (b), (c) and (d) show intensities (or probabilities $p^b$) for the backward-propagating beam from the sources S$_{11}$, S$_{22}$ and S$_{33}$, respectively (red lines). The directions of the incoming beam are denoted by the arrows. 
	Panels (e), (f), (g) show the weak value $(A)_w$ for the case of the forward-propagating beam coming from the source S$_1$ and the backward-propagating beam post-selected by detectors D$_1$, D$_2$ and D$_3$, respectively. Panels (h), (i), (j) show the encounter probability $P=p^fp^b$ for the case of the forward-propagating beam coming from the source S$_1$ and the backward-propagating beam coming from the sources S$_{11}$, S$_{22}$ and S$_{33}$, respectively. Blue lines denote positive and pink lines (in panels (e) and (f)) denote negative value. The magnitude of all quantities is schematically represented by the thickness of the line. The black dashed lines denote the path with zero value of the corresponding quantity. 
}
\end{figure*}

In the TSVF, the forward- and backward-evolving quantum states of light play an important role. For us, this was an inspiration to consider both directions of propagation. In the forward direction (denoted by the index $f$), the intensity distribution of photons (or energy flux density) $I^f_{m,k}$ is proportional to the probability distribution $p^f_{m,k}$
\begin{equation}
\label{If1}
I^f_{m,k}=Kp^f_{m,k}=K|\psi_{m,k}|^2\;,
\end{equation}
which is calculated according to standard quantum theory as the square modulus of the corresponding wave function amplitude $\psi_{m,k}$. $K$ is the proportionality constant; for simplicity we take $K=1$ and so the terms intensity distribution and probability distribution are interchangeable in this paper. As a consequence of the energy conservation law the following sum rule over all optical paths $k$ holds at any stage $m$
\begin{equation}
\label{SumIf2}
\sum_{k=1}^3 p^f_{m,k}=1\;.
\end{equation}

Results of our calculations for the forward direction are schematically shown in 
Fig.~\ref{Fig2}(a). The intensity of the beam at a particular stage of the interferometer is schematically represented by the thickness of the line. Note that in the forward direction in the section between BS$_3$ and BS$_4$, the intensity is zero due to destructive interference and the whole intensity from the nested interferometer propagates towards the detector D$_3$. In order to fully describe the interference including the energy conservation law, we consider detectors in all three quantum channels (D$_1$, D$_2$, and D$_3$). 

For the backward direction (denoted by index $b$), the intensity distribution of photons $I_{m,k,s}^b$ is proportional  to the probability distribution $p^b_{m,k,s}$
\begin{equation}
\label{Ib3}
I_{m,k,s}^b=Kp^b_{m,k,s}=K|\varphi_{m,k,s}|^2\;,
\end{equation}
which is calculated as the square modulus of the corresponding wave function amplitude $\varphi_{m,k,s}$.
In the following, we again choose $K=1$ for simplicity. The index $s=1,2,3$ denotes different experimental configuration using the source S$_{11}$, S$_{22}$ and S$_{33}$, respectively. The intensity is detected at the end of the MZI by the detectors D$_{11}$, D$_{22}$, D$_{33}$. The corresponding intensity distribution is displayed in Figs.~\ref{Fig2}(b), \ref{Fig2}(c) and \ref{Fig2}(d), respectively. Thanks to the energy conservation law, the sum of the probability over all optical paths $k$ yields
\begin{equation}
\label{SumIb4}
\sum_{k=1}^3 p^b_{m,k,s}=1
\end{equation}
for any value of $m$ and $s$. Because of  symmetry reasons, the following sum rule over the experimental configurations $s$ holds as well:
\begin{equation}
\label{SumIbB8}
\sum_{s=1}^3 p^b_{m,k,s}=1
\end{equation}
for any value of $m$ and $k$. 

Next, we confront these results with those of TSVF. 
Following Ref.\cite{Vaidman}, the weak value of the projection operator  $\mathbf{A}_{m,k}$ onto a section with the coordinates $m$ and $k$ is defined as  
\begin{equation}
\label{w5}
(A_{m,k,s})_w=\frac{\mel{\varphi_s}{\mathbf{A}_{m,k}}{\psi}}{\bra{\varphi_s}\ket{\psi}}\;,
\end{equation}
where  $\ket{\psi}$ is the forward-evolving wave function (corresponding to the source S$_1$) and $\bra{\varphi_s}$ is the backward-evolving wave function post-selected by the detector D$_s$. 

Following Li~\etal\cite{Li}, the square modulus of the nominator of Eq. (\ref{w5}) 
\begin{equation}
\label{P6}
P_{m,k,s}=|\mel{\varphi_s}{\mathbf{A}_{m,k}}{\psi}|^2
\end{equation}
is the probability of the detector $s$ clicking under the condition that the photon is found at the section with coordinates $m$ and $k$.  In our paper, we call this quantity an {\it encounter} probability for the reasons that are discussed below. Expressed with the amplitudes of the vectors, Eq.~(\ref{P6}) yields
\begin{equation}
\label{Encounter}
P_{m,k,s}=|\varphi_{m,k,s}|^2\cdot|\psi_{m,k}|^2\;
\end{equation}
and it is obviously equal to the product of probabilities  ($\ref{If1}$) and ($\ref{Ib3}$) of the forward and backward-propagating waves, respectively, 
\begin{equation}
\label{Pb7}
P_{m,k,s}=p^b_{m,k,s}p^f_{m,k}\;.
\end{equation}

Note that within the TSVF framework, $\bra{\varphi_s}$ is the wave function evolving backward in time which is post-selected by the detector D$_s$\cite{Vaidman,Aharonov2007}. In our calculations, this backward-time evolution influences only the results of the weak value~(\ref{w5}). Since the encounter probability (\ref{Encounter}) is given by the square modulus of the wavefunctions, its value is independent of the time direction and it can be thus calculated (and measured) using the beam that is propagating forward in time from sources S$_{ss}$ to detectors D$_{ss}$. 

Using Eq.~(\ref{SumIbB8}) and (\ref{Pb7}), the following sum rules can be derived
\begin{equation}
\label{SumP9}
\sum_{s=1}^3 P_{m,k,s}=p^f_{m,k}\sum_{s=1}^3 p^b_{m,k,s}=p^f_{m,k}
\end{equation}
and using Eq.~(\ref{SumIf2}) we obtain
\begin{equation}
\label{SumSumP10}
\sum_{k=1}^3\sum_{s=1}^3 P_{m,k,s}=1
\end{equation}
for any value of $m$. The latter is a summation rule which is formally similar to the law of conservation of energy, see, e.g., Eq.~(\ref{SumIf2}), however here it is obtained only when the encounter probability $P_{m,k,s}$ is summed up over all experimental arrangements $s$.
For the sake of completeness, we can define normalized encounter probability $\bar{P}_{m,k,s}$ using the Aharonov-Bergmann-Lebowitz rule~\cite{Aharonov2007}
\begin{equation}
\bar{P}_{m,k,s}=\frac{|\mel{\varphi_s}{\mathbf{A}_{m,k}}{\psi}|^2}
{\sum_{k'}|\mel{\varphi_s}{\mathbf{A}_{m,k'}}{\psi}|^2}=\frac{P_{m,k,s}}{\sum_{k'}P_{m,k',s}}
\end{equation}
that can be interpreted as a relative distribution of encounter probability in individual channels.

Numerical results of the weak value (\ref{w5}) for the three variants of $s=1,2,3$ of the backward evolving wavefunction are schematically shown in Figs.~\ref{Fig2}(e), \ref{Fig2}(f) and \ref{Fig2}(g), respectively. Note that the blue and pink lines mean positive and negative value, respectively. Numerical results for the encounter probability $P_{m,k,s}$  are schematically shown in Figs.~\ref{Fig2}(h), \ref{Fig2}(i) and \ref{Fig2}(j), for the three possibilities of the backward-evolving wavefunction, $s=1,2$ and 3, respectively. These patterns can be visually obtained by a ``graphical multiplication" of the intensity pattern of Fig.~\ref{Fig2}(a) with those of Figs.~\ref{Fig2}(b), \ref{Fig2}(c) and \ref{Fig2}(d), respectively.

The interpretation of $P$ (we omit below the indexes $m,k,s$ for brevity) in terms of the multiplication of probabilities $p^f$ and $p^b$ [see Eq.(\ref{Pb7})]  is straightforward. Both probabilities $p^f$ and $p^b$  are given using the standard quantum mechanical interpretation by the square modulus of corresponding amplitudes and meet summation rules~(\ref{SumIf2}) and~(\ref{SumIb4}). Both experiments in the forward and the backward direction are independent, so the product of the probabilities has also meaning of a probability, but rather as a probability of an ``encounter" of photons or alternatively a probability of an independent presence of both forward and backward photons at the same stage at any time. Based on this interpretation, we call $P$  an encounter probability. The encounter probability, see Eq.~(\ref{Pb7}), in contrast with probabilities used in Eqs.~(\ref{If1}), (\ref{Ib3})  does not meet the summation rule similar to Eq.~(\ref{SumIf2}) or~(\ref{SumIb4}). This interpretation of $P$ allows us to understand easily the obtained intensity patterns. In the case of the beam propagating in the forward direction  [see Fig.~\ref{Fig2}(a)], the intensity $I^f$ is zero (and $p^f=0$)  between BS$_3$ and BS$_4$  due to destructive interference. 
In the case of the beam propagating in the backward direction from the sources S$_{11}$ and S$_{22}$,  see Fig. \ref{Fig2}(b) and (c) respectively, the intensity $I^b$ is zero  (and $p^b=0$) between BS$_1$ and BS$_2$ also due to destructive interference. In case of the source S$_{33}$ [see Fig. \ref{Fig2}(d)], the intensity  $I^b$ is zero (and $p^b=0$) in the sections BS$_1$-M$_1$-BS$_4$  and BS$_3$-BS$_4$ for geometric reasons. Consequently, the probability of encounter $P$ given by the product of $p^f$ and $p^b$  is zero in all these sections because of the corresponding reasons.

\begin{figure*}[t]
	\hspace*{-5mm}
	\includegraphics[width=19.5cm]{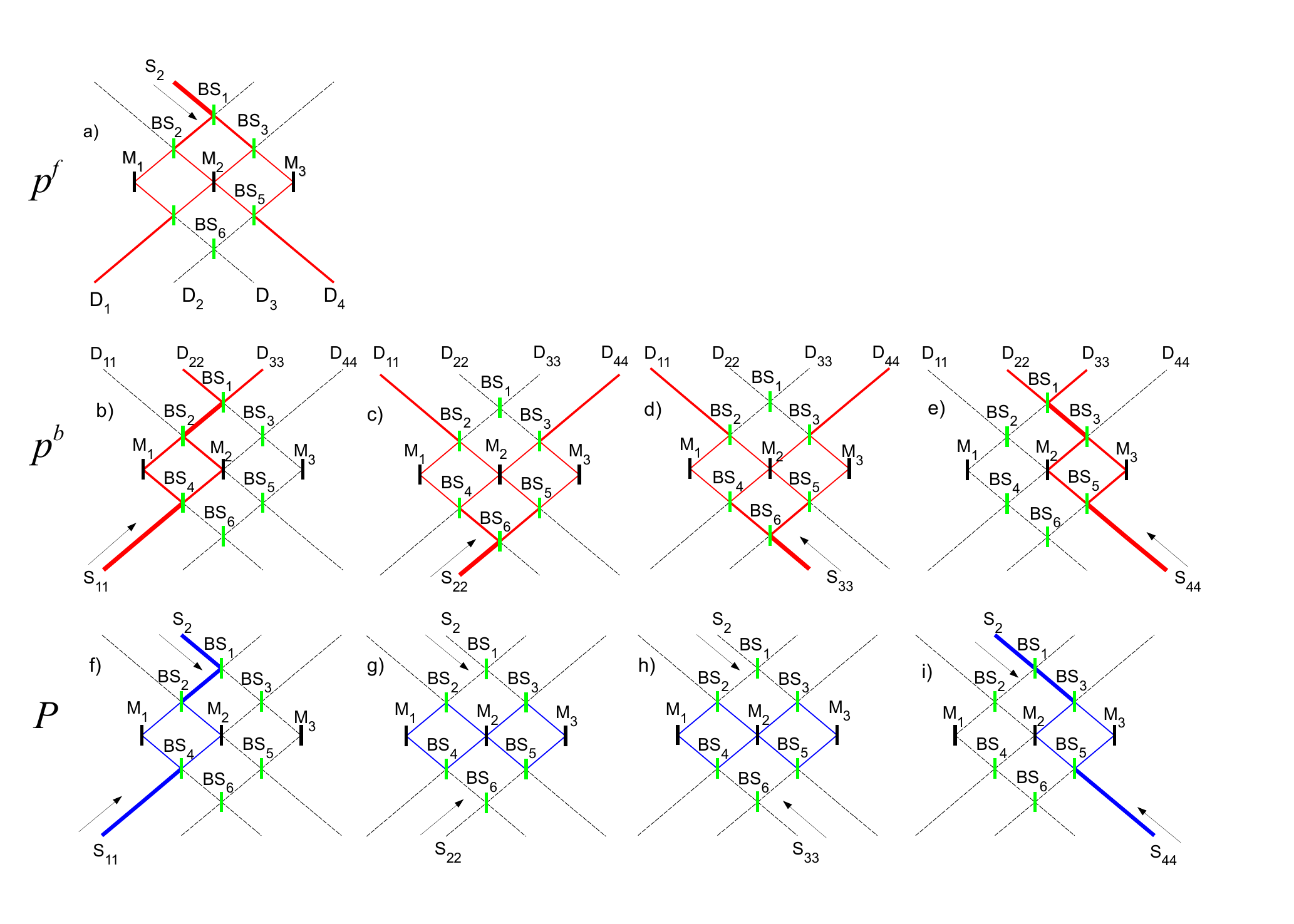}
	\caption{\label{Fig4nnMZ} Results for the double nested  Mach-Zehnder interferometer. 
		M denotes a mirror and BS denotes a beam splitter. Mirror M$_2$ is a mirror reflecting on both sides. Panel (a) schematically shows intensities (or  probabilities $p^f$) of the forward-propagating beam (red lines). Panels (b), (c), (d) and (e) schematically show intensities (or  probabilities $p^b$) of the backward-propagating beam (red lines). The directions of the incoming light are denoted by the arrows.  Panels (f), (g), (h) and (i)  schematically show encounter probability $P=p^fp^b$  (blue lines) for the case of the forward-propagating beam coming from the source S$_1$ and the backward-propagating beam coming from the source  S$_{11}$, S$_{22}$, S$_{33}$ and S$_{44}$, respectively. The magnitude of all quantities is schematically represented by the thickness of the line. The black dashed lines denotes the path with zero value of the corresponding quantity. }
\end{figure*}

The interpretation in terms of TSVF is different. It is claimed that a photon trajectory can be calculated from the forward and the backward-propagating wave functions of the emitted and detected photon and that the weak value is proportional to the trace the photon leaves~\cite{Vaidman,Vaidman2014}. Alternatively, according to  the work of Li~\etal~\cite{Li}, the modulus square of the weak value nominator, see Eq.~(\ref{P6}), represents propability of the detector clicking under the condition that the photon is found at the corresponding section. 
The problems of both of these interpretations clearly shows up in cases displayed in Fig.~\ref{Fig2}(e), \ref{Fig2}(f) and  Fig.~\ref{Fig2}(h), \ref{Fig2}(i), respectively,  where the trajectories in the nested interferometer are completely disconnected from the source and detector since it raises questions, e.g.,  ``how did the photon get to the nested interferometer?" etc. This effect is within the TSVF interpreted as a discontinuity of photon trajectory~\cite{Danan2013}.  However, if it is interpreted as a probability of encounter of photons propagating in the opposite directions, then the discontinuous sections are naturally understood as discussed above.  We do not doubt numerical evaluation of the TSVF calculation reported, e.g., in~\cite{Sokolovski2016,Sokolovski2017,Englert,Griffiths2016,Nikolaev,Duprey}, but we raise questions about the interpretation, legitimacy or physical meaning of the procedure, similarly to discussions and objections of Refs.~\cite{Vaidman2017Comment,Griffiths,VaidmanJETP2017,Nikolaev2017,Sokolovski2017Comment,Peleg, Englert2019}.

\subsection{Double-nested Mach-Zehnder interferometer}
\label{nnMZI}

The single-nested MZI that was discussed above is asymmetric, consequently, the first channel (via mirror M$_1$) is always open. Therefore, in this section we consider a symmetric four-channel interferometer with two nested MZI. We proceed formally in the same way as in Sec.~\ref{nMZI}.

Figure~\ref{Fig4nnMZ}(a) schematically shows the intensity of the beam propagating in the forward direction from  the source S$_2$ to four detectors D$_i$, ($i=1, 2, 3, 4$). Figures~\ref{Fig4nnMZ}(b), \ref{Fig4nnMZ}(c), \ref{Fig4nnMZ}(d) and \ref{Fig4nnMZ}(e) show intensities  of light propagating in the backward direction from the sources $S_{ii}$ to the detectors $D_{ii}$, where $ii=11, 22, 33, 44,$ respectively. The corresponding encounter probabilities calculated using Eq.~(\ref{Pb7}) are presented in Figs.~\ref{Fig4nnMZ}(f), \ref{Fig4nnMZ}(g), \ref{Fig4nnMZ}(h) and \ref{Fig4nnMZ}(i), respectively. Similarly here, this results can be easily visualized by  a ``graphical multiplication" of the corresponding intensity patterns shown in Fig.~\ref{Fig4nnMZ}(a) and Figs.~\ref{Fig4nnMZ}(b), \ref{Fig4nnMZ}(c), \ref{Fig4nnMZ}(d) and \ref{Fig4nnMZ}(e), respectively. In Fig.~\ref{Fig4nnMZ}, we do not show  the weak value because the denominator of Eq.~(\ref{w5}) is zero for this symmetric double nested MZI and the weak value diverges~\cite{Feizpour2011}.

The encounter probabilities shown in  Figs.~\ref{Fig4nnMZ}(f) and \ref{Fig4nnMZ}(i) exhibit continuous   trajectories similarly to standard beam paths, e.g., shown in ~\ref{Fig4nnMZ}(a). In contrast, the encounter probabilities shown in Figs.~\ref{Fig4nnMZ}(g) and \ref{Fig4nnMZ}(h)  are particularly interesting since $P$ is non zero only inside the nested parts of the interferometer. The interpretation of these results in terms of the encounter probability can be easily explained similarly as above using either destructive interference or geometric reasons. However, in terms of the TSVF, the paths are discontinuous without any input nor output and typically the interpretation involves claims about a discontinuous photon trajectory~\cite{Danan2013}. 

\subsection{Simple Mach-Zehnder interferometer}
\label{simpleMZI}
\begin{figure}
	\hspace*{-5mm}
	\includegraphics[width=8cm]{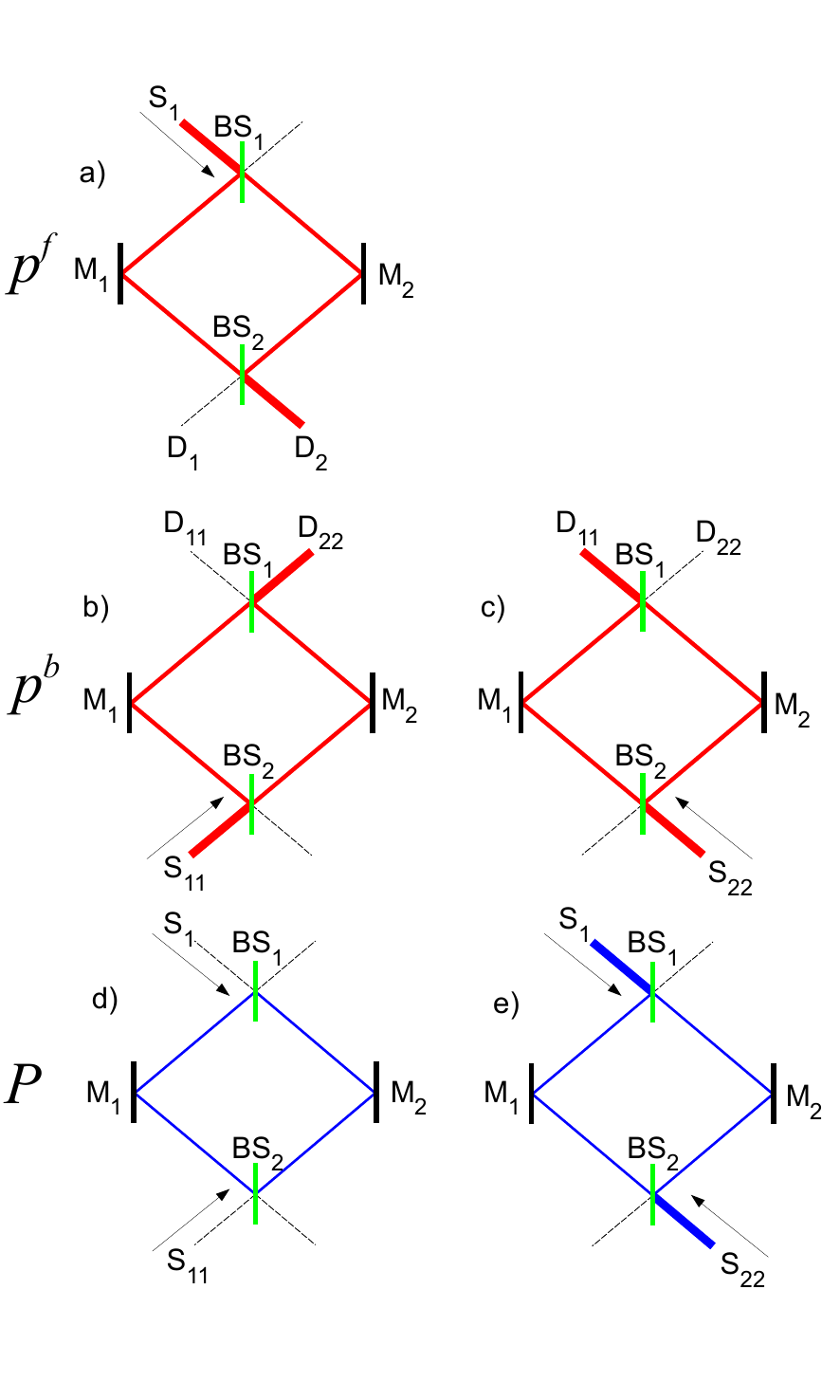}
	\caption{\label{Fig5simpleMZ} Results for the simple Mach-Zehnder interferometer. M denotes mirrors  and BM denotes beam splitters. Panel (a) schematically shows intensities (or  probabilities $p^f$) of the beam propagating in the forward direction (red lines). Panels (b) and (c) schematically show intensities (or  probabilities $p^b$) of the beam propagating in the backward direction (red lines) from the sources S$_{11}$ and S$_{22}$, respectively. The directions of the incoming light are denoted by the arrows. The thickness of the lines on each path is proportional to the intensities of light.
	Panels (d) and (e) schematically show the encounter probability $P=p^fp^b$ (blue lines) on each path for the forward-propagating beam coming from the source S$_1$ and the backward-propagating beam coming from the sources S$_{11}$ and S$_{22}$, respectively. The magnitude of all quantities is schematically represented by the thickness of the line. The black dashed lines denote the path with zero value of the corresponding quantity.
		}
\end{figure} 

In the context of the discussion of the TSVF vs. the encounter probability, it useful to return to the simplest case of the MZI without any nested part that is very often discussed in literature, e.g., in Refs.~\cite{Duarte,Bula,Len}. Using the same procedure as in the case of the single or the double nested MZIs, we calculate the intensities of the forward and the backward-propagating beam and the encounter probability.
Figure~\ref{Fig5simpleMZ}(a) displays the intensity of the forward-propagating beam from the source S$_1$ to the detectors D$_1$ and D$_2$. There is zero signal on the detector D$_1$ due to destructive interference and a unit signal on D$_2$ due to constructive interference. Figures~\ref{Fig5simpleMZ}(b) and \ref{Fig5simpleMZ}(c) display the intensity  of the backward-propagating beam from the sources S$_{11}$ and S$_{22}$, respectively. The detectors D$_{11}$ and D$_{22}$ detect zero signal as shown in Figs.~\ref{Fig5simpleMZ}(b) and \ref{Fig5simpleMZ}(c), respectively, due to destructive interference. Figures~\ref{Fig5simpleMZ}(d) and \ref{Fig5simpleMZ}(e) display the encounter probability, that can be as above visualized as the product of the probabilities shown in Fig.~\ref{Fig5simpleMZ}(a) and Figs.~\ref{Fig5simpleMZ}(b), \ref{Fig5simpleMZ}(c), respectively. 
Note that again the denominator of the weak value, see Eq.~(\ref{w5}), is zero for this symmetric MZI and the weak value diverges~\cite{Feizpour2011}.

\begin{figure*}
	\includegraphics[width=13cm]{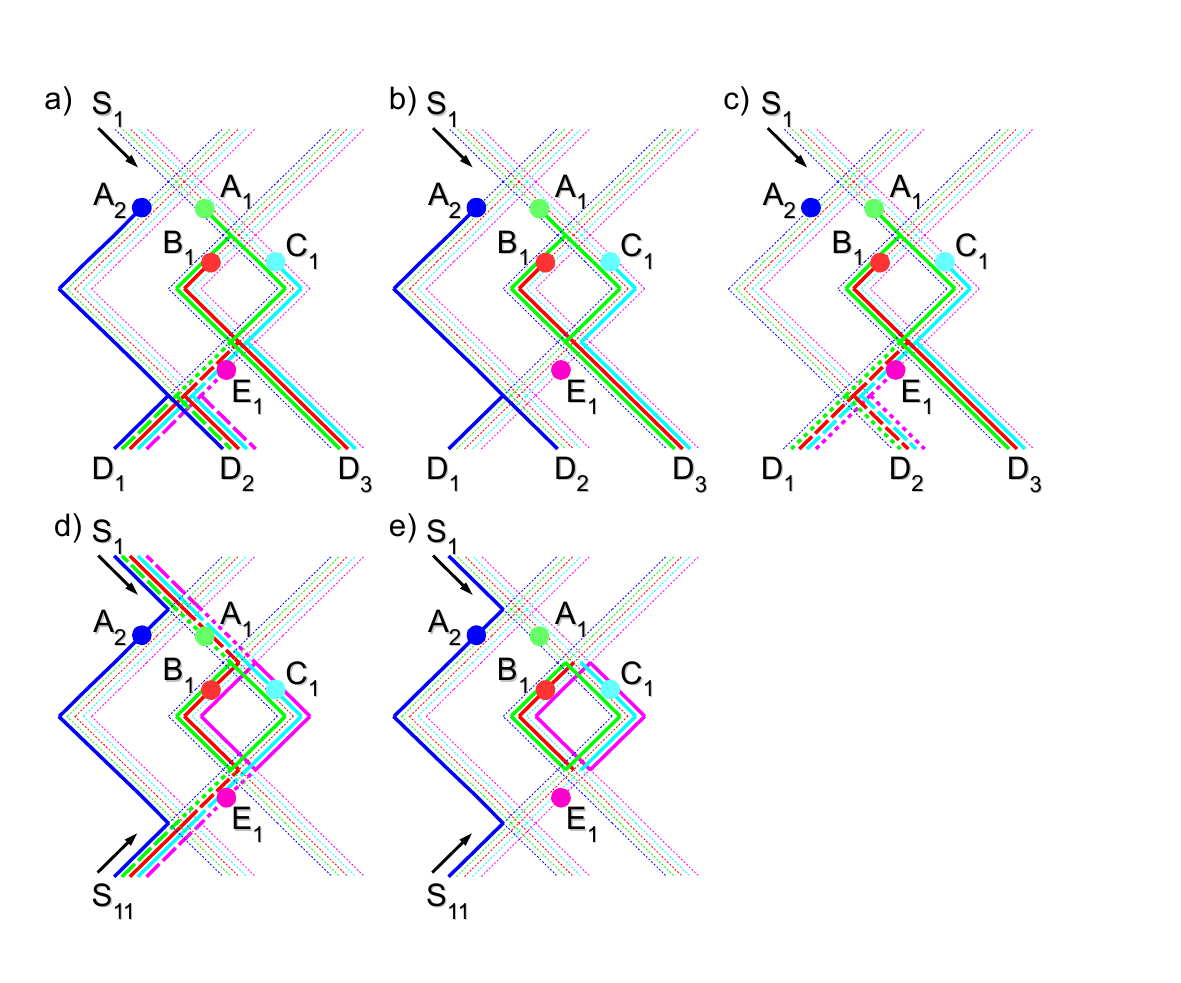}
	\vspace*{-10mm}
	\caption{\label{Fig5perturb} Results for the nested Mach-Zehnder interferometer  with active modulators  A$_2$, A$_1$, B$_1$, C$_1$ and E$_1$ denoted by blue, green, red, cyan and magenta circles, respectively. Some symbols shown in Fig.~\ref{Fig1nMZscheme} are omitted for clarity. The thick lines denote Fourier perturbation amplitudes of intensity (or the encounter probability) with color corresponding to the modulators. The order of perturbation is differentiated by the line type (solid line for first order, dashed line for the second order and dotted line for the third order). Thin dotted lines represents the signals with zero Fourier amplitudes. Panel (a) displays the Fourier perturbation amplitudes of the beam propagating in the forward direction from the source S$_1$ when all modulators work at different frequencies, i.e., are independent. Panel (b) displays the same as (a) except that modulators B$_1$ and C$_1$ are identical. Panel (c) displays the situation of panel (a) with modulator A$_2$ blocking the beam.
		Panels (d) and (e) display	the Fourier perturbation amplitudes of the encounter probability for the forward beam shown in (a) and (b), respectively, with the backward-propagating beam coming from the source S$_{11}$.}
\end{figure*}

According to the TSVF, Figs.~\ref{Fig5simpleMZ}(d) and \ref{Fig5simpleMZ}(e) show a trajectory of photons when the photon was detected by D$_1$ or D$_2$, respectively. Again here we see a similar situation as above where some of the patterns exhibit continuous trajectories, see Fig.~\ref{Fig5simpleMZ}(e), but the other, see Fig.~\ref{Fig5simpleMZ}(d), shows zero probability neither on input nor on the output of the interferometer, however, there is still a non-zero probability inside of the interferometer. The interpretation of $P$ shown in Fig.~\ref{Fig5simpleMZ}(d) in terms of the encounter probability involves simple arguments about destructive or constructive interference, however, the interpretation in terms of the TSVF has to involve speculations about a discontinuous photon trajectory~\cite{Danan2013}. Already this most simple MZI exhibits the key properties that occur in the more complex nested MZIs discussed above. 

\subsection{The nested Mach-Zehnder interferometer with perturbations}

\label{perturbations}
We believe that the issue is essentially explained above. However, since perturbations of MZI are often used in experiments, we devote this section to  the discussion of this topic. Consider modulators A$_1$, A$_2$, B$_1$, C$_1$, E$_1$ shown in Fig.~\ref{Fig1nMZscheme}. Inspired by the modulation spectroscopy~\cite{Cardona}, we prefer modulators based on the change of amplitude or phase of the beam rather than a change of the beam direction. For simplicity we assume transmission modulators that modulate the amplitude and in principle the phase. The transmission coefficient of the modulator for a wave function (or plane wave) is assumed in the form 
\begin{equation}
\label{tau11}
\tau_X=[\tau_{0X}-\epsilon_{0X}\cos(2\pi f_Xt)]{\rm e}^{\rm i\delta_X}
\end{equation}
where $\tau_{0X}$  is the unperturbed transmission coefficient (for simplicity we choose $\tau_{0X}=1$),   $\epsilon_{0X}\ll1$ is the perturbation amplitude (we assume a real number), $f_X$  is the frequency of the modulation and  $\delta_X$ is a phase shift. For an easy comparison with the literature we use zero phase shifts  $\delta_X=0$, however the calculations can be easily extended for non-zero phase shifts; see, e.g.,  Ref.~\cite{Potocek}. 

One experimentally often used way of distinguishing the influence of different modulators in the detected signal is by using different modulation frequencies $f_X$; see, e.g., Ref.~\cite{Danan2013}.  Consequently we examine the Fourier transform of the detected signal and we take the Fourier coefficient with the corresponding frequency as a measure of the detected perturbation amplitude, see also Refs.~\cite{Huang,Wu,Wiesniak2018}. In the following, we call it the perturbation amplitude. We describe in detail here only the single-nested MZI; for the double-nested and simple MZIs, the conclusions are analogous. We choose the frequencies of the modulators A$_2$, A$_1$, B$_1$, C$_1$, E$_1$ in the following order: 3, 5, 7, 11 and 17 Hz. The choice of frequencies is not essential; however it is advisable to avoid overlaps in their mutual combinations.  All modulators work simultaneously.

The results for the perturbation amplitude in the forward direction are schematically displayed in Fig.~\ref{Fig5perturb}(a). For the sake of visibility, the same line thickness is used for all signals with non-zero intensity and the order of perturbation in $\epsilon_{0X}$ is differentiated by the line type (solid line for first order, dashed line for the second order and dotted line for the third order). The total signal is a function of frequency with the first order signals having the fundamental frequencies. The frequency of the second  and third order signals is given by the combinations of their corresponding fundamental frequencies. 
For simplicity, we display the signal only with the lowest order from a given modulator.

Figure~\ref{Fig5perturb}(a) depicts that the detector D$_1$ detects signal from the modulators A$_2$, B$_1$, C$_1$ in the first order and A$_1$, E$_1$ in the second order; the same situation is on the detector D$_2$. The latter was already described in work of M.~Wiesniak~\cite{Wiesniak2018}.
The detector D$_3$ detects only the first order signals from A$_1$, B$_1$, and C$_1$. 
The section of the optical path with the modulator E$_1$ (between BS$_3$ and BS$_4$) is particularly interesting. Recall that it exhibits the complete destructive interference in the case of the unperturbed MZI, see Fig.~\ref{Fig2}(a). However, in the case of the perturbed MZI, there is a signal but only in the second order from the  modulators  B$_1$, C$_1$ and in the  third order from the  modulators A$_1$, E$_1$ depicting the imperfect destructive interference.

If the transmission coefficients $\tau_{\rm B1}$ and $\tau_{\rm C1}$  are equal, the complete destructive interference in the section with modulator E$_1$ is restored. 
This situation is displayed in Fig.~\ref{Fig5perturb}(b) and depicts that there are neither signals A$_1$, B$_1$, C$_1$, E$_1$ on the detectors D$_1$ and D$_2$ nor on the optical path in the section with the modulator E$_1$. These results are obvious in terms of classical optics and standard quantum theory. 

In the work of Danan~\etal\cite{Danan2013} they reported the case when the optical path at the modulator A$_2$ is blocked and no modulated signal on the detector D$_2$ was observed,  see Fig. 2(c) in Ref.~\cite{Danan2013}. In our case this situation would correspond to $\tau_{A2}=0$ and the corresponding intensity in the forward direction is shown in Fig.~\ref{Fig5perturb}(c).  We see that the signals from A$_1$, B$_1$, C$_1$, E$_1$ can be detected on the detectors D$_1$ and D$_2$ but only in the second and the third order of the detected signal. This signal was probably not resolved in Ref.~\cite{Danan2013} because the higher order signals have much lower intensity. Similar conclusion (the absence of detection of higher order signals) can be made for the similar experimental work of Zhou~\etal\cite{Zhou} that used the single photon source.

So far in the case of perturbations we calculated only the intensities of the forward propagated beam. Formally we can calculate also the backward-propagating wave and calculate the encounter probability $P$ using Eq.~(\ref{Pb7}). The results for 
the encounter probability calculated for perturbations shown in Figs.~\ref{Fig5perturb}(a) and \ref{Fig5perturb}(b) and for the backward beam with $s=1$, (i.e., coming from the source S$_{11}$) are shown in Fig.~\ref{Fig5perturb}(d) and \ref{Fig5perturb}(e), respectively. We can see that the encounter probability corresponding to the case of the imperfect destructive interference, see Fig.~\ref{Fig5perturb}(d), looks continuous, at  least in some higher order signals in perturbation, however, in case of the perfect destructive interference, see Fig.~\ref{Fig5perturb}(e), it exhibits discontinuous sections. Analogous conclusions can be obtained for the double nested or simple MZI.

\section{Summary}
We have presented calculations of the intensity of the beam propagating in a nested, a double nested and a simple Mach-Zehnder interferometer calculated using classical optics and standard quantum theory. Qualitatively, the results can be understood as a result of constructive or destructive interference. We show that the probability of detection of photon derived from the weak value used in the TSVF formalism can be interpreted as the probability of encounter of two opposing photon fluxes. This interpretation does not need to involve any discontinuous photon trajectories often used in TSVF interpretations. We discussed also the perturbations of the nested Mach-Zehnder  interferometer and showed that the signals from the modulators can propagate in the case of the imperfect destructive interference in the second or the third order in perturbation. 
\begin{acknowledgments}
We acknowledge helpful discussions with D. Munzar and J. Rusna\v{c}ko.
This work was supported by the MEYS of the Czech Republic under the project CEITEC 2020 (LQ1601). 
\end{acknowledgments}

\end{document}